\documentclass{article}
\usepackage[preprint]{neurips_2026}
\usepackage[utf8]{inputenc}
\usepackage[T1]{fontenc}
\usepackage{hyperref}
\usepackage{xurl}
\usepackage{booktabs}
\usepackage{amsfonts}
\usepackage{amsmath}
\usepackage{microtype}
\usepackage{graphicx}
\usepackage{float}

\title{Black--Scholes for the land:\\ modelling savannah response to rain as a dynamic world model}

\author{Tom Pitts\\ University of Technology Sydney\\ \texttt{thomas.pitts@uts.edu.au}}

\begin{document}
\maketitle

\begin{abstract}
Carbon credits, offsets and index insurance rest on counterfactuals: what the land would have done under the same weather without management, or what a bad season cost. Current baselines fail in two ways: some ignore rain, others condition on rain linearly and annually. But the true response lags the rain, depends on hidden state such as moisture from other sources, and responds differently from pixel to pixel. Focussing on grassland pasture as a proof of concept, we fit a temporal world model of MODIS greenness conditioned on CHIRPS rain over two $45$km$^2$ sites in Laikipia, Kenya: a transition density map for greenness given recent and current rainfall. We use it, conceptually, as Black--Scholes is used. Inverting the model on observed greenness it yields an implied residual, everything the land did that the rain does not explain. Two capabilities drop out of this solution: 1) attribution, what the land did beyond the rain; 2) sensitivities, derivatives of the rollout with respect to rain and state, close analogues of the Greeks. On Laikipia test sites, our learned model recovers three-month greenness with $23\%$ less error than a history-only baseline and $13\%$ less than the best hand-written model on unseen years. In the 2022 drought, the land ended $0.17$ NDVI below normal; the rain that failed explains roughly half of that, and under normal rain with the same residual it would still have ended $0.10$ below. $\Delta$, the derivative of greenness with respect to rain, is computed at every pixel and averages $0.075$ NDVI per extra $100$\,mm of rain. A $\Delta$ at every pixel is an exposure to rain that can be hedged, allowing efficient pricing of contracts and accurate attribution.
\end{abstract}

\section{Approach}
Various methodologies exist to measure the carbon capture attributable to sustainable grassland management practices, such as rotation of grazing animals, replanting strategies, and backburning. The Northern Kenya Rangelands Carbon Project, the world's largest soil carbon project, earns one credit per tonne of soil carbon above a baseline set by VM0032, a pioneering grazing methodology \citep{verra2023}. That baseline was contested enough for Verra to suspend issuance for eight months in 2023 \citep{verra2023,si2023}. The methodology never asks how the land responds to the rain. Elsewhere, livestock index insurance in the same counties pays on seasonal NDVI against the pixel's own history \citep{ibli}. For maximum efficiency, the insurer wants forage lost to drought, not to overgrazing. Both payments need what the land did split into the rain's share and the rest, and to the best of our knowledge, no operational baseline makes that split cleanly. The pixel's history incorporates temporal weather effects as land management. In our model, we estimate the specific terrain response from data, as a transition density $p_\theta(a_{t+1}\mid a_t, w_t, r_{t+1})$ for the greenness anomaly $a$ given a rain-memory state $w$ and rainfall $r$.

A Black-Scholes formula (BSF) model \cite{bs} is fed observables such as current stock price, strike price, and used to derive a parameter value -- implied volatility -- which cannot be observed directly but is implied by the market's current pricing of options. In our model the observables are the total state and the rain that fell, and the model is inverted on observed greenness. What comes out is the residual, what the land did beyond what the rain explains. Furthermore, specific values can also be plugged back in to the model. The model can be re-run with the residual held fixed and the rain replaced, and it returns the land under that other rain with everything else as it was. Derivatives come by automatic differentiation, as the Greeks come from the BSF.

\paragraph{Residual} The residual between modelled and observed includes everything not in the model. That includes grazing, clearing and fire, the management and disturbance signal itself, but also error in the CHIRPS rainfall product, error in the fitted response, and satellite noise. We assume error in CHIRPS and noise are statistically constant over a sufficiently large dataset.

\section{Setup}
\paragraph{Data} The target $a_t$ is greenness, the normalised difference vegetation index (NDVI) from the MODIS satellite product MOD13Q1, at $250$\,m every 16 days (one composite), 2002--2025, as an anomaly against each pixel's 2003--2018 climatology. The z-score used throughout is this anomaly divided by the pixel's standard deviation for that time of year. The input $r_t$ is rain from CHIRPS summed over the same 16-day windows, in units of $100$\,mm. The setting is two $45$\,km boxes in Laikipia, Kenya, with rains in March--May and October--December: Box A (development) and Box B (held-out landscape) to its west (Figure~\ref{fig:map}, Appendix). We fit on Box A 2003--2017, early-stop on 2018, and evaluate on Box A 2019--2025, 14 seasons including the 2021--22 drought and 2023 recovery, and on Box B in every year.

\paragraph{Model} Each step sees only the current state, not the series i.e. Markovian. The state is the anomaly $a_t$ and an antecedent-rain index $w_t$, a one-parameter bucket whose memory $\lambda(x)$ is fitted per pixel:
\begin{equation}
w_{t+1} = \lambda(x)\,w_t + r_{t+1}, \qquad
a_{t+1} = a_t + \mu_\theta(a_t, \bar a_t, w_t, r_{t+1}, \phi_{t+1}, x) + \sigma_\theta(\cdot)\,\varepsilon_t,\quad \varepsilon_t\sim\mathcal N(0,1),
\label{eq:model}
\end{equation}
with $\bar a_t$ the $5\times5$ neighbourhood mean and $\phi$ the seasonal phase. \textbf{M0} uses history alone. \textbf{M1} has $\mu$ and $\log\sigma$ linear in the features. \textbf{M1.5} adds a hand-written S-curve in $w_t+r_{t+1}$, six parameters. \textbf{M2} adds a 64-unit one-hidden-layer network to $\mu$ and $\log\sigma$. All are fitted by full-batch Adam on the one-step Gaussian log-likelihood of observed transitions.

\paragraph{Residual} Inverting \eqref{eq:model} on an observed transition gives the implied innovation,
\begin{equation}
\hat\varepsilon_t = \frac{a_{t+1} - a_t - \mu_\theta(a_t,\bar a_t,w_t,r_{t+1},\phi_{t+1},x)}{\sigma_\theta(\cdot)},
\label{eq:implied}
\end{equation}
one number per pixel per 16 days, in units of the model's own spread. Accumulated over a season it is the attribution product: $a_{t+h}-F_h$, where $F_h=\mathbb E[a_{t+h}]$ is the mean rollout along the rain that fell, with the model's spread as its interval. The counterfactual re-runs \eqref{eq:model} from the same state along another rain path $r'$ with $\hat\varepsilon$ held fixed:
\begin{equation}
\tilde a_{t+1} = \tilde a_t + \mu_\theta(\tilde a_t,\bar{\tilde a}_t,\tilde w_t,r'_{t+1},\phi_{t+1},x) + \sigma_\theta(\cdot)\,\hat\varepsilon_t,\qquad \tilde w_{t+1}=\lambda(x)\,\tilde w_t + r'_{t+1},\qquad \tilde a_t = a_t.
\label{eq:cf}
\end{equation}
This is the abduction, action, prediction counterfactual of structural causal models: infer the exogenous term, change the input, roll forward. Any rain path can be chosen for $r'$. If set to the rain that fell, \eqref{eq:cf} returns the observation exactly; if set to the average rain for that time of year, it returns the land under normal rain with residual intact. Sensitivities (following BSF Greeks) are Jacobian-vector products through the rollout: $\Delta_h=\partial F_h/\partial r_{t+1}$, greenness $h$ steps ahead per $100$\,mm of rain now; $\Theta_h=\partial F_h/\partial a_t$, how much of today's anomaly survives. The anomaly score is $-\log p_\theta(a_{t+1}\mid a_t,w_t,r_{t+1})$ i.e. the surprise of the rain-conditioned model at what the satellite saw. N.B. the residual and the score are (approximately) two readings of the same quantity.

\paragraph{Evaluation} Error is the continuous ranked probability score (CRPS), the probabilistic generalisation of absolute error, on observed held-out pixel-steps (one pixel at one 16-day composite); skill is $1-\mathrm{CRPS}/\mathrm{CRPS}_{\mathrm{M0}}$. Each model is scored one step (16 days) ahead, and rolled forward three months along the rain that fell.

\section{Results}
\begin{figure}[t]
\centering\includegraphics[width=\linewidth]{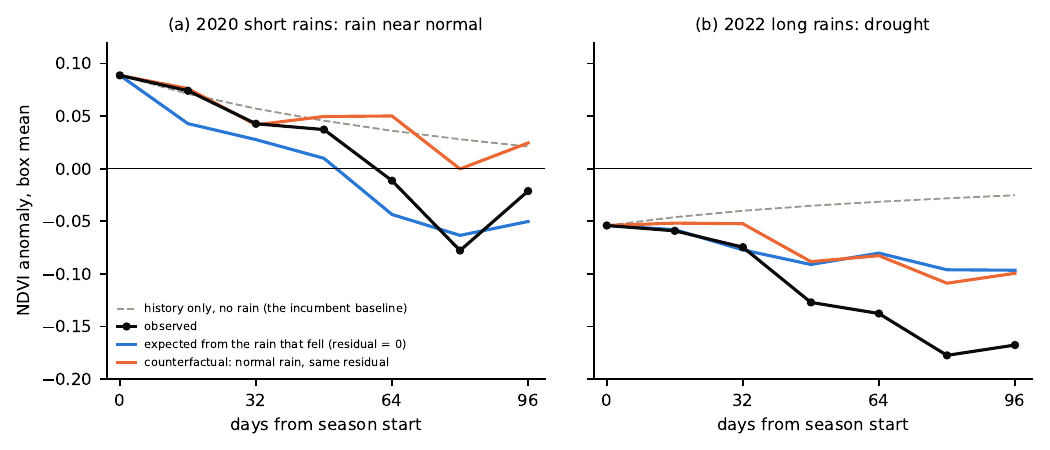}
\caption{The residual backed out and plugged in, Box A, held-out years, box means. Black: observed. Grey dashed: history only, the incumbents' baseline. Blue: what the learned model expects from the rain that fell, residual set to zero. Orange: the counterfactual of \eqref{eq:cf}, normal rain with the observed residual re-injected. Observed minus blue is the residual; observed minus orange is what the rain did. (a) Near-normal rain: the model tracks the land and there is little to correct. (b) The 2022 drought: history says recovery, the model says half the fall, the land fell further.}
\label{fig:residual}
\end{figure}

\paragraph{Fit} M2's one-step skill exceeds M1's by $0.07$ on unseen years and $0.05$ on the unseen landscape, and its spread is calibrated (Appendix). Along the rain that fell, three-month error is $23\%$ below history and $13\%$ below the S-curve on unseen years, $17\%$ and $8\%$ on the unseen landscape (Figure~\ref{fig:response}, Table~\ref{tab:onestep}). Fed climatological rain the same network falls behind history within two months; fitted without rain its skill is $0.016$. The S-curve's amplitude fitted near zero: the nonlinearity is in how rain interacts with pixel, state and season, not in rain alone.

\paragraph{Residual and counterfactual} Figure~\ref{fig:residual}. In October--December 2020 rain was $38$\,mm below normal and the counterfactual ends within $0.05$ NDVI of the observation: near-normal rain, near-zero rain effect. In March--May 2022 rain was $124$\,mm below normal: the box ended $0.17$ NDVI below normal. History alone predicts recovery, so a z-score assigns the whole $0.17$ to the land. The model assigns $0.07$ to the rain that failed, $0.07$ to the residual, and the rest to the deficit the season began with; under normal rain with the residual held fixed the land ends $0.10$ below. The orange line is the baseline for a credit; observed minus orange is the rain's share, the quantity an index insurer pays on. Where rain is more than $60$\,mm below normal the residual is systematically negative: part of the 2022 residual is model under-response to a drought deeper than its training. Read as a likelihood, the same residual tells drought from disturbance: with every score tuned to flag $5\%$ of unburned land, the two drought seasons draw $24\%$ false alarms from a z-score and $5\%$ from the model (Table~\ref{tab:anomaly}). Scoring the residual against mapped burns is out of scope here; protocol, figure and full results are in the Appendix (Figure~\ref{fig:fire}).

\paragraph{Sensitivities} 
\begin{table}[t]\centering\footnotesize\setlength{\tabcolsep}{3pt}
\caption{Does $\Delta$ rank patches by their real response to rain? Held-out rainy-season steps 2019--25, in thirds by M2's $\Delta$ at the starting state, split at each box's median rain departure. Response: NDVI change two steps on per $100$\,mm of rain anomaly, given starting greenness; $\pm$ two bootstrap s.d.; bold clears zero.}\label{tab:greeks}
\begin{tabular}{llccccc}\toprule
 & Rain & \multicolumn{3}{c}{Real response, by $\Delta$ third} & Model $\Delta$ & Most $-$\\
Landscape & departure & least & medium & most & by third & least\\\midrule
A, unseen years & $\le 30$\,mm & $.096\pm.092$ & $.088\pm.100$ & $.147\pm.096$ & $.041/.087/.124$ & $.051\pm.160$\\
 & $> 30$\,mm & $.034\pm.018$ & $.066\pm.034$ & $.078\pm.038$ & $.032/.088/.123$ & $\mathbf{.044\pm.040}$\\
B, unseen land & $\le 21$\,mm & $.083\pm.066$ & $.286\pm.238$ & $.237\pm.140$ & $.064/.097/.122$ & $\mathbf{.154\pm.135}$\\
 & $> 21$\,mm & $.062\pm.042$ & $.087\pm.048$ & $.131\pm.034$ & $.066/.098/.123$ & $\mathbf{.070\pm.051}$\\\bottomrule
\end{tabular}
\end{table}

\textbf{$\Delta$ -- } On the grid mean an extra $100$\,mm of rain in March--May raises greenness by a (mean) peak of $0.075$ NDVI a month later, positive in sign at every pixel and half gone by three months ($.029/.075/.073/.060/.047/.036$ by 16-day step). Recomputed from the current state at each rainy-season step, $\Delta$ sorts exposure well. Sorted by $\Delta$, the top third of patches green about twice as much per $100$\,mm as the least, on unseen years and landscape, with the right magnitudes on that landscape (Table~\ref{tab:greeks}). We also find that the response is a property of the pixel's state, not of the pixel: a frozen map averaged over the training years is no better than a single constant (Table~\ref{tab:greekstests}), and its rank correlation with the real response drops from $0.70$ to $0.38$ frozen on Box A.

\textbf{$\Theta$ -- } A measure of impact decay over time, the $\Theta$ produced here predicts the impact of exogenous shock halves in $35$ days in the long rains and $38$ in the dry, against $38$ days in the anomaly's raw autocorrelation (Figure~\ref{fig:greeks}). Black--Scholes $\Theta$ describes how a position decays while nothing happens; ours is the rate an anomaly decays with the rain path held fixed and no new residual. In other words, we align naming for the role rather than the precise derivative.

\textbf{$\partial\sigma/\partial r$ -- } The model predicts both a spread $\sigma$ and a mean $\mu$. An extra $100$\,mm widens the one-step spread by a mean average of $21\%$ on Box A and $32\%$ on the unseen landscape, at every pixel individually. Held-out steps bear it out: sorted by the rain that fell, predicted spread rises from $.060$ to $.082$ as realised error rises from $.061$ to $.093$, although the model in its current configuration still under-predicts spread through the middle of that range. Black--Scholes has no equivalent because it assumes $\sigma$ constant -- the volatility smile \citep{rubinstein1994,dupire1994} is an artifact of this assumption's limitation.

\section{Conclusion}
Inverting our learned transition density world model on observed satellite data produces the residual, what the land did beyond the rain's effect. By differentiation, it returns key sensitivities --- what a pixel's greenness is attributable to, how long a shock to it lasts, and how much rain widens the range of outcomes. Read as a likelihood, it tells drought from exogenous disturbance. In March--May 2022, a held-out year, Box A ended $0.17$ NDVI below normal. The model accounts $0.07$ to the rain that failed, and re-run under normal rain with the residual held fixed the land still ends $0.10$ below, representing a baseline a credit or a payout should be measured against (Figure~\ref{fig:residual}). A z-score would predict recovery that season and attribute the whole $0.17$ to management. While only a proof of concept on limited data and with various simplifications, we believe this invertible transition density approach has real potential to practical world modelling of complex real-world temporal systems that has genuine utility.

\appendix
\section{Location, data, tables, calibration, fitting}
\begin{figure}[H]
\centering\includegraphics[width=0.6\linewidth]{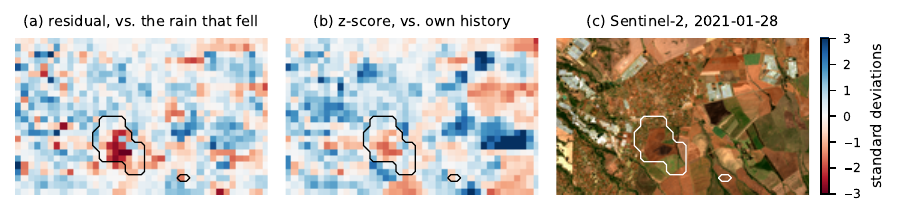}
\caption{A burn on 17 January 2021 (outlined, 50 pixels), scored one composite later. (a) The residual of \eqref{eq:implied}: the burn is the red patch. (b) The z-score, the same composite against the pixel's own history: the burn is invisible. (c) Sentinel-2 true colour ten days after the burn: a harvested field on irrigated cropland, $80\%$ of the outline by area.}
\label{fig:fire}
\end{figure}

\paragraph{Data and preparation} Three sources are fused onto one grid and one calendar. The surface is MODIS Terra MOD13Q1 v061 NDVI at $250$\,m, one composite per 16 days from 2002 to 2025 (537 composites), warped to a $0.0025^\circ$ grid of $160\times160$ pixels per box. Pixels whose reliability flag exceeds 1 (cloud, snow, fill; $2\%$ of pixel-steps) are treated as unobserved: their values are interpolated in time so the model always has an input state, but they are never used as targets. Each pixel's climatology is its mean NDVI per composite-of-year over 2003--2018, and the anomaly $a_t$ is NDVI minus that climatology. The driver is CHIRPS v2 daily rainfall at $0.05^\circ$, summed over each composite's 16-day window and bilinearly resampled to the grid, in units of $100$\,mm, with a climatology built the same way. Static covariates $x$ are the pixel's mean NDVI, seasonal amplitude and climatological annual rain. MCD64A1 burned area gives the day each $500$\,m cell burned, and a pixel is marked burned in the composite window containing that day. The MODIS record has gaps: no composites for 61 days after 1 November 2021, so October--December 2021, one of the two drought seasons, contributes two composites, and about 30 days after 2 February 2023, 27 July 2024 and 2 December 2024. Steps spanning a gap are not yet excluded from any analysis.
\paragraph{Residual runs} The residual and counterfactual of Figures~\ref{fig:residual} and~\ref{fig:fire} use the fitted M2 with no refitting; each season starts from the observed state before its first 16-day window and runs six windows. Re-injecting $\hat\varepsilon$ along the rain that fell reproduces the observed anomaly to a maximum absolute error of $10^{-6}$ in every season. Box-mean anomaly at three months for all 13 held-out seasons is in the supplementary run directory. The year-to-year persistence of the residual and its forecast value were tested with a per-pixel, per-year latent in the drift, inferred by MAP on one year and used to forecast the next: the latent beats a pixel-shuffled control on both boxes but improves three-month error by $0.3$--$0.6\%$, not significant on either. Residual against rain, held-out rainy-season pixel-steps: correlation $0.05$ ($0.03$ on training years); mean residual in spread units by rain departure, below $-60$\,mm / $-60$ to $-30$ / $-30$ to $0$ / $0$ to $30$ / $30$ to $60$ / $60$ to $120$ / above $120$\,mm: $-0.24$, $-0.03$, $-0.03$, $+0.22$, $+0.05$, $+0.10$, $-0.01$ on held-out years against $-0.04$, $+0.07$, $+0.12$, $+0.16$, $+0.13$, $+0.09$, $+0.20$ on training years. Fire, 17 January 2021 burn scored on the 2 February composite: mean residual on burned pixels $-0.75$ against $-0.04$ elsewhere; mean z-score $-0.17$ against $-0.13$.
\begin{figure}[H]
\centering\includegraphics[width=\linewidth]{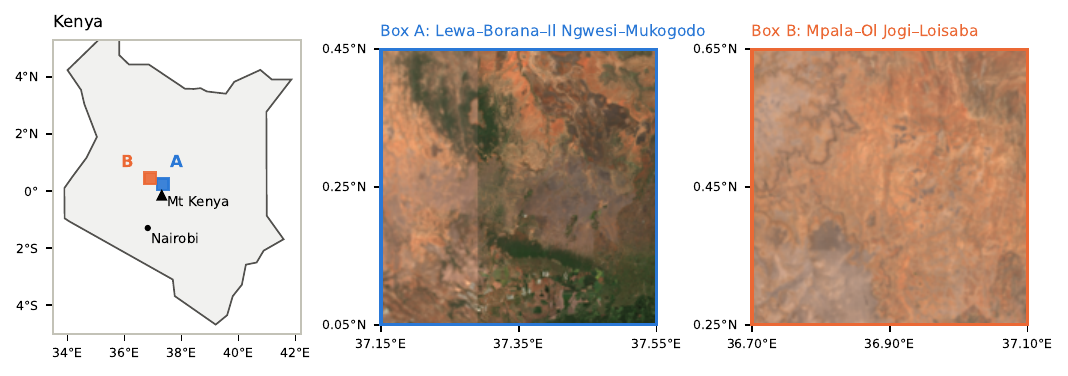}
\caption{Where. Left: Kenya with the two $45$\,km boxes; Box A spans $37.15$--$37.55^\circ$E, $0.05$--$0.45^\circ$N, Box B $36.70$--$37.10^\circ$E, $0.25$--$0.65^\circ$N. Centre and right: Sentinel-2 true colour of each box in the 2023 dry season; dark green is forest and riverine woodland, pale is open rangeland.}
\label{fig:map}
\end{figure}
\begin{table}[H]\centering\caption{One-step scores on observed held-out transitions. NLL is Gaussian negative log-likelihood per transition; skill is $1-\mathrm{CRPS}/\mathrm{CRPS}_{\mathrm{M0}}$.}\label{tab:onestep}\small
\begin{tabular}{lcccc}\toprule
 & \multicolumn{2}{c}{Box A, 2019--25} & \multicolumn{2}{c}{Box B, all years}\\
Model & NLL & CRPS skill & NLL & CRPS skill\\\midrule
M0 history only & -1.208 & 0.000 & -1.448 & 0.000\\
M1 linear, realised rain & -1.227 & 0.013 & -1.543 & 0.052\\
M1.5 S-curve, realised rain & -1.226 & 0.009 & -1.543 & 0.052\\
M2 without driver & -1.277 & 0.016 & -1.504 & 0.023\\
M2 learned, realised rain & -1.277 & 0.085 & -1.628 & 0.104\\
\bottomrule\end{tabular}
\end{table}
\paragraph{Anomaly protocol} A pixel-step is flagged when its score exceeds a cutoff. The three scores (M2, M1, z-score) are on different scales, so each gets its own cutoff, set to flag $5\%$ of pixel-steps where nothing burned. With false alarms equalised, we ask what fraction of burned pixels each score flags within a month of the burn, and what fraction of pixel-steps it flags in the two drought seasons, when the land browned but nothing burned. This is a probe rather than a benchmark: one landscape, 208 burned pixel-composites, a score that flags only drops --- a variant adopted after the pre-registered two-sided score was run --- and one of the two drought seasons, October--December 2021, reduced to two composites by a gap in the MODIS record.
\begin{table}[H]\centering\caption{Anomaly detection scored on fire, Box A 2019--25, 208 burned pixel-composites. Hit rate: burns flagged within two composites. Lag: mean composites from burn to flag among hits. FAR: unburned pixel-steps flagged in October--December 2021 and March--May 2022. AUC: rank separation of burns from unburned pixel-steps. Upper block: pre-registered two-sided score; lower block: one-sided drop score, post hoc.}\label{tab:anomaly}\small
\begin{tabular}{llcccc}\toprule
 & Score & hit rate @5\% FAR & lag (steps) & FAR, drought seasons & AUC\\\midrule
two-sided, pre-reg. & z-score & 0.06 & 0.54 & 0.017 & 0.558\\
 & M1 & 0.19 & 0.23 & 0.060 & 0.713\\
 & M1.5 & 0.19 & 0.23 & 0.061 & 0.710\\
 & M2 & 0.19 & 0.28 & 0.018 & 0.735\\
\midrule
one-sided, post hoc & z-score & 0.15 & 0.45 & 0.244 & 0.755\\
 & M1 & 0.27 & 0.25 & 0.148 & 0.711\\
 & M1.5 & 0.27 & 0.25 & 0.146 & 0.705\\
 & M2 & 0.28 & 0.24 & 0.051 & 0.773\\
\bottomrule\end{tabular}
\end{table}
\begin{figure}[H]
\centering\includegraphics[width=0.7\linewidth]{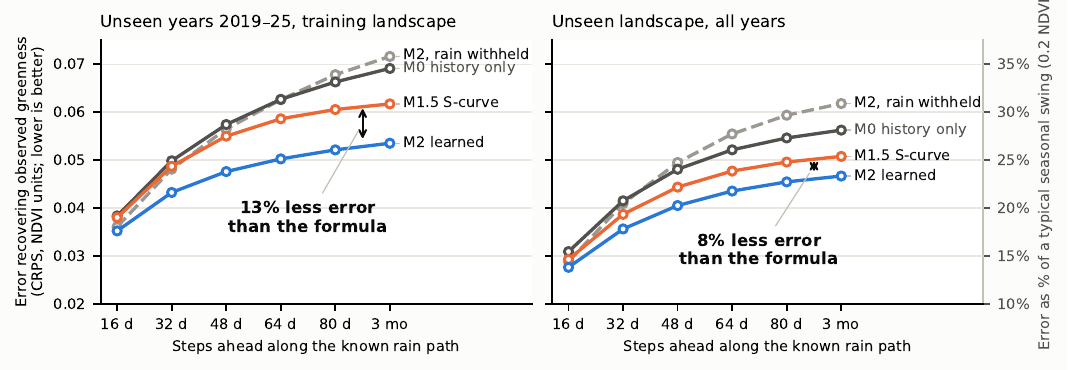}
\caption{Given the rain that fell, how well does each model recover what the land did? CRPS of sampled rollouts against the observed anomaly along the known rain path; no model uses a rain forecast. Grey dashed: M2 fed climatological rain. Right axis: error as a share of a typical $0.2$ NDVI seasonal swing.}
\label{fig:response}
\end{figure}
\begin{figure}[H]
\centering\includegraphics[width=0.6\linewidth]{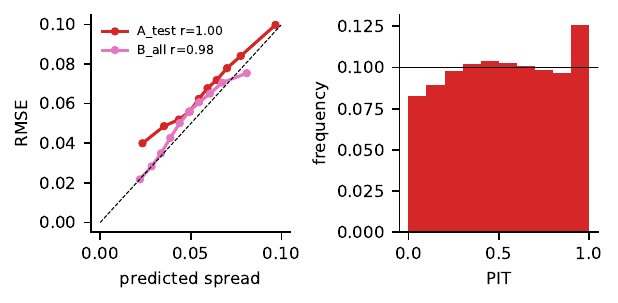}
\caption{M2 calibration on held-out sets: RMSE against predicted spread by decile, and the PIT histogram on Box A 2019--25.}
\end{figure}
\begin{figure}[H]
\centering\includegraphics[width=0.96\linewidth]{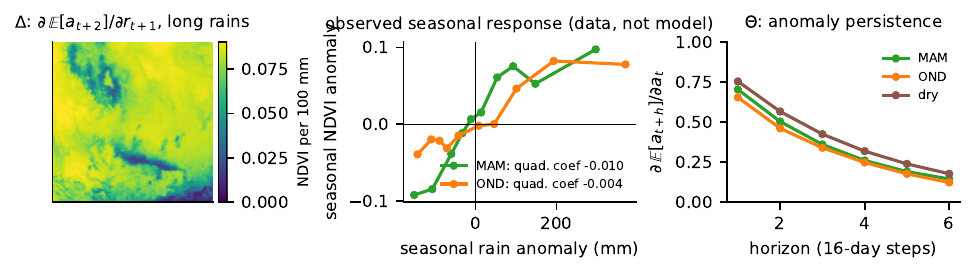}
\caption{Box A, held-out years. Left: $\Delta$ of the fitted density two steps after $100$\,mm of extra rain in the long rains. Middle: the observed seasonal response, binned NDVI anomaly against seasonal rain anomaly, with its quadratic coefficient; a property of the data, not a model derivative. Right: $\Theta=\partial F_h/\partial a_t$, persistence of an anomaly by horizon and season. Its half-life is $35$ days in the long rains, $34$ in the short and $38$ in the dry, against $38$ days in the anomaly's own $16$-day autocorrelation over 2003--2018 ($.73/.56/.30/.17$ at $1/2/4/6$ steps). That is a model-free check, and one $\Theta$ should undercut slightly, since it holds the rain memory $w$ fixed and so drops the persistence that is really rain's: M2 does ($.70/.50/.26/.15$), while the S-curve runs long in the other direction ($.79/.62/.38/.24$, $46$ days).}
\label{fig:greeks}
\end{figure}
\begin{table}[H]\centering\caption{Pre-registered tests of $\Delta$, written after fitting and before running; primary G1 on Box A. Gaps are real response of the most minus the least $\Delta$ third (NDVI per $100$\,mm); G3 is two-step skill against persistence of $\alpha+\beta\,\Delta\,x$. Brackets: $95\%$ bootstrap interval over test years.}\label{tab:greekstests}\small
\begin{tabular}{lcc}\toprule
Test & Box A & Box B\\\midrule
G1 (primary): within pixel & $+.026$ $[-.026, +.070]$ fail & $+.064$ $[+.014, +.116]$\\
G2: across pixels, all rain & $+.047$ $[+.005, +.084]$ pass & $+.062$ $[+.009, +.110]$\\
G3: skill, minus constant & $+.010$ $[-.023, +.050]$ fail & $+.006$ $[-.005, +.018]$\\
G3: skill, minus frozen map & $+.002$ $[-.031, +.044]$ fail & $+.007$ $[-.004, +.019]$\\
Size: small minus large & $+.008$ $[-.102, +.176]$ fail & $+.084$ $[-.060, +.221]$ fail\\
M1.5, across pixels & $-.052$ $[-.122, -.001]$ & $+.008$ $[-.117, +.085]$\\\bottomrule
\end{tabular}\end{table}
\paragraph{Fitting} We use full-batch Adam at a learning rate of $3\times10^{-3}$, running 800 steps for M0 and M1 and 2000 for M1.5 and M2, with early stopping on 2018 log-likelihood; each step uses all $25\,600$ pixels and the $\sim340$ training-year transitions. M2 fits in 190\,s on one H100, and a refit moves one-step skill by about $0.005$. The fitted rain memory $\lambda$ is $0.16$ for M1, $0.21$ for M1.5 and $0.38$ for M2. Jacobian-vector products of the six-step mean rollout agree with central finite differences to $10^{-4}$ absolute, against a $\Delta$ of order $10^{-2}$. Code and the run directory behind every number are in the supplementary repository.

\begin{thebibliography}{9}\raggedright
\bibitem[Seddon et al.(2016)]{seddon2016} A. W. R. Seddon, M. Macias-Fauria, P. R. Long, D. Benz, K. J. Willis. Sensitivity of global terrestrial ecosystems to climate variability. \emph{Nature} 531, 2016.
\bibitem[Funk et al.(2015)]{chirps} C. Funk et al. The climate hazards infrared precipitation with stations. \emph{Scientific Data} 2, 2015.
\bibitem[Didan(2021)]{mod13} K. Didan. MODIS/Terra Vegetation Indices 16-Day L3 Global 250m SIN Grid V061. NASA EOSDIS LP DAAC, 2021.
\bibitem[Giglio et al.(2018)]{mcd64} L. Giglio, L. Boschetti, D. P. Roy, M. L. Humber, C. O. Justice. The Collection 6 MODIS burned area mapping algorithm and product. \emph{Remote Sensing of Environment} 217, 2018.
\bibitem[Verra(2023)]{verra2023} Verra. Update: Northern Kenya Grassland Carbon Project. Program notice, 26 March 2023. \url{https://verra.org/program-notice/update-northern-kenya-grassland-carbon-project/}
\bibitem[Survival International(2023)]{si2023} Survival International. Verra's `review' of the Northern Rangelands Trust carbon project. November 2023. \url{https://assets.survivalinternational.org/documents/2501/2311_Response_to_Verra_review_of_NRT_carbon_project.pdf}
\bibitem[Chantarat et al.(2013)]{ibli} S. Chantarat, A. G. Mude, C. B. Barrett, M. R. Carter. Designing index-based livestock insurance for managing asset risk in northern Kenya. \emph{Journal of Risk and Insurance} 80, 2013.
\bibitem[Black and Scholes(1973)]{bs} F. Black, M. Scholes. The pricing of options and corporate liabilities. \emph{Journal of Political Economy} 81, 1973.
\bibitem[Rubinstein(1994)]{rubinstein1994} M. Rubinstein. Implied binomial trees. \emph{Journal of Finance} 49(3), 1994.
\bibitem[Dupire(1994)]{dupire1994} B. Dupire. Pricing with a smile. \emph{Risk} 7(1), 1994.
\end{thebibliography}
\end{document}